\documentclass[journal,onecolumn,12pt]{IEEEtran}
\usepackage{cite}
\ifCLASSINFOpdf
   \usepackage[pdftex]{graphicx}
   \DeclareGraphicsExtensions{.pdf,.jpeg,.png}
\else
   \usepackage[dvips]{graphicx}
   \DeclareGraphicsExtensions{.eps}
\fi
\usepackage[cmex10]{amsmath}
\usepackage{amssymb}
\usepackage{amsbsy}
\usepackage{algorithmicx,algorithm}
\usepackage{algpseudocode}
\usepackage{enumerate}
\usepackage{threeparttable}

\newtheorem{sec2_lemma1}{Lemma}[section]
\newtheorem{sec2_lemma2}[sec2_lemma1]{Lemma}
\newtheorem{sec2_lemma3}[sec2_lemma1]{Lemma}
\newtheorem{sec2_lemma4}[sec2_lemma1]{Lemma}
\newtheorem{sec2_lemma5}[sec2_lemma1]{Lemma}
\newtheorem{sec2_lemma6}[sec2_lemma1]{Lemma}
\newtheorem{sec2_lemma7}[sec2_lemma1]{Lemma}
\newtheorem{sec2_lemma8}[sec2_lemma1]{Lemma}

\newtheorem{sec3_thm1}{Theorem}[section]
\newtheorem{sec3_thm2}[sec3_thm1]{Theorem}

\newtheorem{sec3_lemma1}{Lemma}[section]
\newtheorem{sec3_lemma2}[sec3_lemma1]{Lemma}

\begin{document}
\title{On Some Exponential Sums Related to the Coulter's Polynomial}

%\author{Minglong~Qi~\IEEEmembership{Member,~IEEE,
%} Shengwu~Xiong, Jingling~Yuan, Wenbi~Rao, and Luo~Zhong
\author{Minglong~Qi, Shengwu~Xiong, Jingling~Yuan, Wenbi~Rao, and Luo~Zhong
     % <-this % stops a space
\thanks{The authors are with the School of Computer Science and Technology, Wuhan University of Technology, Mafangshan West Campus, 430070 Wuhan, China (e-mail: mlqiecully@163.com)}% <-this % stops a space
%\thanks{Minglong Qi was supported by the Foundation for the project \textit{Secure Middleware Technologies} of  Wuhan University of Technology under Grant 2012037.  Shengwu~Xiong and Jingling~Yuan were supported by the Foundation for the project \textit{Formation Model for Excellent Engineers in Software Engineering} of Hubei Province under Grant 2012103.}
}

\maketitle

\begin{abstract}
%Linear codes with few weights have many applications ranging from consuming electronics,  massive storage, information security, to combinatorial mathematics. In this letter, some new linear codes with two or three weights are constructed based on the Coulter's work on exponential sums over finite field, and by using the genetic construction method called the defining set method firstly introduced by Ding et el, and their weight distribution are settled as well. 
In this paper, the formulas of some exponential sums over finite field, related to the Coulter's polynomial, are settled based on the Coulter's theorems on Weil sums, which may have potential application in the construction of linear codes with few weights.
\end{abstract}

\begin{IEEEkeywords}
 Gauss sum, exponential sum over finite field, the Coulter's polynomial, linear code, set cardinality.
\end{IEEEkeywords}

% \fi
\IEEEpeerreviewmaketitle

\section{Introduction}
\IEEEPARstart{T}{hroughout}  this paper, 
%let $ p $ be an odd prime and $ q=p^{e} $ with $ e$ a positive integer. Two symbols are kept unchanged as well: $  \alpha $ being a positive integer and $ d=\gcd(\alpha,e)$. In this paper, it is always supposed that $ \frac{e}{d} $ is odd. 
we fix the following notations:
\begin{itemize}
\item $ p $ is an odd prime, and $ q=p^{e} $ with $ e$ a positive integer.
\item $  \alpha $ is a positive integer, and $ d=\gcd(\alpha,e)$.
\item $ \mathbb{F}_{q} $ denotes the Galois field with $ q $ elements, whose cyclic multiplicative group is  $ \mathbb{F}_{q}^{*} $.
\item $ \overline{f(x)} $ is the complex conjugate of the function $ f(x) $ over $ \mathbb{F}_{q} $, while $ \widehat{f}(x) $ is the same function but with $ x\in \mathbb{F}_{p}  $.
\item $ Tr(x) $ is the absolute trace mapping from $ \mathbb{F}_{q} $ onto  $ \mathbb{F}_{p} $. 
\item $ \zeta_{p} =e^{\frac{2\pi\sqrt{-1}}{p}}$.
\end{itemize}

Let $ a\in \mathbb{F}_{q}^{*} $.
$ \mathfrak{C}(x)=a^{p^{\alpha}}x^{p^{2\alpha}} +ax \in \mathbb{F}_{q}[x] $ 
 is called the Coulter's polynomial over $ \mathbb{F}_{q} $, based on which is or not a permutation polynomial of $ \mathbb{F}_{q} $, and on how whose solution set is for a given $ b\in \mathbb{F}_{q}^{*} $ such taht $ \mathfrak{C}(x)=-b^{p^{\alpha}} $, Coulter gave the evaluation of two exponential sums of type Weil \cite{Bib9,Bib10},  which have  important applications in the coding theory, information security, and combinatorics, for instance see \cite{Bib6,Bib8,Bib11,Bib12,Bib13}.
 
Let $ a,c\in \mathbb{F}_{p} $, and $ b\in \mathbb{F}_{q}^{*} $. The two exponential sums related to the Coulter's polynomial, and to be studied in this paper, are defined below.
\begin{align}
A_{\alpha}(a)&=\sum_{y\in \mathbb{F}_{p}^{*}}\zeta_{p}^{-ay}\sum_{x\in \mathbb{F}_{q}}\zeta_{p}^{yTr(x^{p^{\alpha}+1})}\label{s1_Aa_definition},\\
B_{\alpha}(a,c)&=\sum_{y\in \mathbb{F}_{p}^{*}}\sum_{z\in \mathbb{F}_{p}^{*}}\zeta_{p}^{-ay-cz}\sum_{x\in \mathbb{F}_{q}}\zeta_{p}^{Tr(yx^{p^{\alpha}+1}+zbx)}\label{s1_Bac_definition}.
\end{align}
 
 The algebraic evaluation of (\ref{s1_Aa_definition}) and (\ref{s1_Bac_definition}) is crucial in the determination of the cardinalities of some subsets of $ \mathbb{F}_{q} $, which are involved in the construction of linear codes with few weights, especially by using the generic method called \textit{defining set} method, firstly introduced by Ding et al \cite{Bib1,Bib3}, and much studied later, see, for instance \cite{Bib5,Bib6,Bib7,Bib8}. In this paper, we are concerned with the following two subsets of $ \mathbb{F}_{q} $:
 
 \begin{align}
 D_{\alpha}(a)&=\lbrace x\in \mathbb{F}_{q} : Tr(x^{p^{\alpha}+1})=a \rbrace,\label{s1_definingset}\\
 M_{\alpha}(a,c)&=\lbrace x\in \mathbb{F}_{q} : Tr(x^{p^{\alpha}+1})=a\ \text{and}\ Tr(bx)=c \rbrace.\label{s1_definition_Mac}
 \end{align}
 
 The subset (\ref{s1_definingset}) was used in \cite{Bib8} as the defining set for case $ \frac{e}{d} $ being even. If both $ a=0 $ and $ c=0 $, the defining set (\ref{s1_definingset}) may be made smaller by a factor $ p-1 $,  to derive the punctured version of the original linear codes \cite{Bib3,Bib6}.
 
 The cardinalities of (\ref{s1_definingset}) and (\ref{s1_definition_Mac}) are defined below:
 \begin{equation}\label{codewords_length}
       n_{\alpha}(a)=
       \begin{cases}
       | D_{\alpha}(a) \cup \lbrace 0\rbrace |, &\qquad\text{if}\ a=0,\\
       | D_{\alpha}(a)|,&\qquad\text{if}\ a\ne 0,
       \end{cases}
       \end{equation}
       and
       \begin{equation}\label{Mac_length}
        N_{\alpha}(a,c)=|M_{\alpha}(a,c)|.
       \end{equation}
       
 In this paper, it is aimed at establishing the algebraic formulas to the exponential sums (\ref{s1_Aa_definition}) and (\ref{s1_Bac_definition}), and then applying them to evaluate the cardinalities of the subsets (\ref{s1_definingset}) and (\ref{s1_definition_Mac}), i.e., $ n_{\alpha}(a) $ and $ N_{\alpha}(a,c) $, for case $ \frac{e}{d} $ being odd. 
% The main tools are the Coulter's formulae on Weil sums and their refinements  \cite{Bib8,Bib9,Bib10}. 
 The rest of the paper is organized as follows: in Section II, mathematical background such as group characters, Gauss sum, and the Coulter's formulae, is presented. The formulae for the  exponential sums (\ref{s1_Aa_definition}) and (\ref{s1_Bac_definition}), and for the set cardinalities (\ref{s1_definingset}) and (\ref{s1_definition_Mac}), are given in Section III, along with their proofs. Finally, a brief concluding remarks are given in Section IV.

\section{Mathematical background}
  
%   This section is devoted to prove the main theorems of the previous section, before that some concepts and basic results on group character and a series of needed lemmas, such as the work of Coulter on exponential sum over finite field \cite{Bib9,Bib10}, some intermediate results in \cite{Bib7}, will be presented.
   
   An \textit{additive character} of $ \mathbb{F}_{q} $, $ \chi $, is a nonzero function from $ \mathbb{F}_{q} $ to a set of nonzero complex numbers such that for any pair $ (x,y) \in \mathbb{F}_{q}\times \mathbb{F}_{q} $, $ \chi (x+y) =\chi(x)\chi(y)$. In this paper, the complex conjugate of $ \chi $ is denoted by $\overline{\chi} $. For each $ b\in \mathbb{F}_{q}$, an additive character of  $ \mathbb{F}_{q} $ can be defined below
   \begin{equation}\label{additive_character}
   \chi_{b}(c)=\zeta_{p}^{Tr(bc)},\ \text{for all}\ c\in \mathbb{F}_{q},
   \end{equation}
   where $ \zeta_{p}=e^{\frac{2\pi\sqrt{-1}}{p}} $. In (\ref{additive_character}), when setting $ b=0 $, the resultant character $ \chi_{0} $ is said to be \textit{trivial} since for all $  c\in \mathbb{F}_{q} $, $ \chi_{0}(c) =1$. The character $ \chi_{1} $ is called the \textit{canonical additive character}. It was shown that  any additive character of 
  $ \mathbb{F}_{q} $ can be written as $ \chi_{b}(x)=\chi_{1}(bx) $ \cite[Chapter 5]{Bib15}. In this paper, the canonical additive character is used and its subscript omitted.
  
   The orthogonal property of the additive character over $ \mathbb{F}_{q} $ is resumed in the following \cite[Chapter 5]{Bib15}:
  \begin{equation*}
  \sum_{x\in \mathbb{F}_{q}}\chi(bx)=
  \begin{cases}
  q,&\ \text{if}\ b=0,\\
  0,&\ \text{otherwise.}
  \end{cases}
  \end{equation*}
  
  A \textit{multiplicative character} $ \psi $ of $ \mathbb{F}_{q} $ over $ \mathbb{F}_{q}^{*} $ is a nonzero function from $ \mathbb{F}_{q}^{*} $ to  a set of nonzero complex number such that for any $ (x,y)\in \mathbb{F}_{q}^{*}\times\mathbb{F}_{q}^{*}$,  $ \psi(xy)=\psi(x)\psi(y) $. Let $ \theta $ be a primitive element of $ \mathbb{F}_{q}^{*} $. Then, any multiplicative character over $ \mathbb{F}_{q}^{*} $ can be written as 
  \begin{equation*}
  \psi_{j} (\theta^{k}) =e^{2\pi\sqrt{-1} jk/(q-1)},    
  \end{equation*}
  where $ 0\leq j,k\leq q-2 $. The multiplicative character $ \psi_{(q-1)/2} $ is called \textit{the quadratic character} of $ \mathbb{F}_{q} $, denoted by $ \eta $. In this paper, suppose that $ \eta(0)=0 $. With the canonical additive character and the quadratic character,   \textit{the quadratic Gauss sum} over $ \mathbb{F}_{q} $ can be defined by
  \begin{equation}\label{Gauss_sum_def}
  G(\eta,\chi)=\sum_{x\in \mathbb{F}_{q}^{*}}\eta(x)\chi(x).
  \end{equation}
  
  In order not to confuse with the complex conjugation, let $ \widehat{\chi} $ denote the canonical additive character over $ \mathbb{F}_{p} $, and $ \widehat{\eta} $ the quadratic character over $ \mathbb{F}_{p} $, respectively. 
%  Let $ \bigl(\frac{t}{p}\bigr) $ denote the Legendre symbol of $ t $ modulo $ p $.
  
  \begin{sec2_lemma1}[\cite{Bib15}, Theorem 5.15]\label{s2_lemma1_gauss_sum}
  Let the symbols be that presented before. Then,
  \begin{equation*}
  \begin{cases}
  G(\eta,\chi)&=(-1)^{e-1}\sqrt{-1}^{\frac{(p-1)^{2}e}{4}}\sqrt{q},\\
  G(\widehat{\eta},\widehat{\chi})&=\sqrt{-1}^{\frac{(p-1)^{2}}{4}}\sqrt{p}.
  \end{cases}
  \end{equation*}
  \end{sec2_lemma1}
  
  In some situation, it is necessary to know what is the value of $ \eta(x) $ when $ x\in \mathbb{F}_{p}^{*} $, which is answered by the following lemma \cite[Lemma 7]{Bib6}:
  
  \begin{sec2_lemma2}\label{s2_lemma2_quaChar}
  Let the symbols be that presented before. Then,
  \begin{equation*}
  \eta(x) =
  \begin{cases}
  1,&\qquad\text{if}\  e\ \text{is even,}\\
  \widehat{\eta}(x),&\qquad\text{if}\  e\ \text{is odd,}
  \end{cases}        
  \end{equation*}
  where $ x\in \mathbb{F}_{p}^{*} $.
  \end{sec2_lemma2}
  
  Next lemma \cite[Theorem 5.33]{Bib15} gives the explicit evaluation of exponential sum of the quadratic polynomial over finite field, $ f(x)=a_{2}x^{2}+a_{1}x+a_{0}\in \mathbb{F}_{q}[x] $, with $ a_{2} \ne 0$:
  
  \begin{sec2_lemma3}\label{s2_lemma2_quafunc}
  \begin{equation*}
  \sum_{x\in \mathbb{F}_{q}}\chi(f(x))=\chi(a_{0}-a_{1}^{2}(4a_{2})^{-1})\eta(a_{2})G(\eta,\chi).
  \end{equation*}
  \end{sec2_lemma3}
  
  Recall that $ q=p^{e} $ where $ p $ is an odd prime and $ e $ a positive integer, $ d=\gcd(e,\alpha) $. 
%  and that $ \frac{e}{d} $ is supposed always to be odd. 
  Let $ \mathfrak{C}(x)=a^{p^{\alpha}} x^{p^{2\alpha}}+ax\in \mathbb{F}_{q}[x]$. We call $  \mathfrak{C}(x) $ the Coulter's polynomial, based on which is or not a permutation polynomial over $ \mathbb{F}_{q} $, Coulter  \cite{Bib10} completely determined the following Weil sum:
  \begin{equation}\label{s2_coulter_sab}
  S_{\alpha}(a,b)= \sum_{x\in \mathbb{F}_{q}}\chi(ax^{p^{\alpha}+1}+bx),
  \end{equation}
  where $ a\in \mathbb{F}_{q}^{*}, b\in \mathbb{F}_{q} $.
  
%  Let $ a\in \mathbb{F}_{q}^{\times} $ and $ b\in  \mathbb{F}_{q} $ where $ q=p^{e} $ with $ p $ an odd prime and $ e $ a positive integer. In \cite{Bib10}, Coulter gave the definition of a kind of exponential sum as below
%  \begin{equation}\label{Coulter_sum_sab}
%  S_{\alpha}(a,b)=\sum_{x\in\mathbb{F}_{q}}\chi(ax^{p^{\alpha}+1}+bx),
%  \end{equation}

  The explicit formulae to evaluate (\ref{s2_coulter_sab})  are resumed in the next two lemmas:
  
  \begin{sec2_lemma4}[\cite{Bib10},Theorem 1]\label{s2_lemma4_sab_perm}
  Let $ q $ be odd and suppose $ \mathfrak{C}(x) $ is a permutation polynomial over $ \mathbb{F}_{q} $. Let $ x_{0} $ be the unique solution of the equation $ \mathfrak{C}(x)=- b^{p^{\alpha}},b\ne 0$. The evaluation of (\ref{s2_coulter_sab}) partitions into the following two cases.
  \begin{enumerate}[(1)]
  \item If $ e/d  $ is odd then
  \begin{equation}\label{formula_sab_edodd}
   S_{\alpha}(a,b)=
   \begin{cases}
   (-1)^{e-1}\sqrt{q}\eta(-a)\overline{\chi(ax_{0}^{p^{\alpha}+1}})&\qquad\text{if}\ p\equiv 1\pmod 4\\
   (-1)^{e-1}\sqrt{-1}^{3e}\sqrt{q}\eta(-a)\overline{\chi(ax_{0}^{p^{\alpha}+1}})&\qquad\text{if}\ p\equiv 3\pmod 4.
    \end{cases}
  \end{equation}
   \item If $ e/d  $ is even then $ e=2m $, $ a^{(q-1)/(p^{d}+1)}\ne (-1)^{m/d} $ and
     \begin{equation*}
     S_{\alpha}(a,b)=-(-1)^{m/d}p^{m}\overline{\chi(ax_{0}^{p^{\alpha}+1}}).
     \end{equation*}
  \end{enumerate}
  \end{sec2_lemma4}
  
  \begin{sec2_lemma5}[\cite{Bib10},Theorem 2]\label{s2_lemma5_sab_nonperm}
  Let $ q=p^{e} $ be odd and suppose $ \mathfrak{C}(x) $ is not a permutation polynomial over $ \mathbb{F}_{q} $. Then for $ b\ne 0 $ we have $ S_{\alpha}(a,b)=0 $ unless the equation $ \mathfrak{C}(x)=- b^{p^{\alpha}}$ is solvable. If this equation is solvable, with some solution $ x_{0} $ say, then
  \begin{equation*}
  S_{\alpha}(a,b)=-(-1)^{m/d}p^{m+d}\overline{\chi(ax_{0}^{p^{\alpha}+1}}).
  \end{equation*}
  \end{sec2_lemma5}
  
  To ease symbol manipulation, in this paper we define the following constant, relative to the formulae to evaluate (\ref{s2_coulter_sab}) when $ \frac{e}{d} $ is odd:
  \begin{equation}\label{s2_coulter_constant}
  \kappa=
  \begin{cases}
  (-1)^{e-1}\sqrt{q}, &\qquad\text{if}\ p\equiv 1\pmod 4,\\
  (-1)^{e-1}\sqrt{-1}^{3e}\sqrt{q}, &\qquad\text{if}\ p\equiv 3\pmod 4.
  \end{cases}
  \end{equation}
  
  The following lemma that computes the constant in (\ref{s2_coulter_constant}) is straightforward:
  \begin{sec2_lemma6}\label{s2_lemma6_coulter_cst}
  Let the symbols be that of (\ref{s2_coulter_constant}) and before, then
  \begin{itemize}
  \item If $ e=2m $, then
  \begin{equation}\label{s2_coulter_cst_eeven}
    \kappa=
    \begin{cases}
    -p^{m}, &\qquad\text{if}\ p\equiv 1\pmod 4,\\
    -(-1)^{m}p^{m}, &\qquad\text{if}\ p\equiv 3\pmod 4.
    \end{cases}
    \end{equation}
   \item If $ e $ is odd, then
     \begin{equation}\label{s2_coulter_cst_eodd}
       \kappa=
       \begin{cases}
       p^{\frac{e}{2}}, &\qquad\text{if}\ p\equiv 1\pmod 4,\\
       -\sqrt{-1}^{e}p^{\frac{e}{2}}, &\qquad\text{if}\ p\equiv 3\pmod 4.
       \end{cases}
       \end{equation}
  \end{itemize}
  \end{sec2_lemma6}
  
  A special case of (\ref{s2_coulter_sab}), $ S_{\alpha}(a,0) $  where $a\in \mathbb{F}_{q}^{*} $, was studied in \cite{Bib9}, resumed in the following two lemmas:
  
%  \begin{sec2_lemma7)\label{s2_lemma7_sa0_edodd}
%  Let $ \frac{e}{d} $ be odd. Then,
%  \begin{equation*}
%     S_{\alpha}(a,0)=
%     \begin{cases}
%     (-1)^{e-1}\sqrt{q}\eta(a), &\qquad\text{if}\ p\equiv 1\pmod 4\\
%     (-1)^{e-1}\sqrt{-1}^{e}\sqrt{q}\eta(a),&\qquad\text{if}\ p\equiv 3\pmod 4.
%      \end{cases}
%    \end{equation*}
%  \end{sec2_lemma7}
\begin{sec2_lemma7}[\cite{Bib9}, Theorem 1]\label{s2_lemma7_sa0_edodd}
Let $ \frac{e}{d} $ be odd. Then,
\begin{equation*}
   S_{\alpha}(a,0)=
   \begin{cases}
   (-1)^{e-1}\sqrt{q}\eta(a),&\qquad\text{if}\ p\equiv 1\pmod 4\\
   (-1)^{e-1}\sqrt{-1}^{e}\sqrt{q}\eta(a),&\qquad\text{if}\ p\equiv 3\pmod 4.
    \end{cases}
  \end{equation*}
\end{sec2_lemma7}  

\begin{sec2_lemma8}[\cite{Bib9}, Theorem 2]\label{s2_lemma8_sa0_edeven}
Let $ \frac{e}{d} $ be even with $ e=2m $. Then,
\begin{equation*}
   S_{\alpha}(a,0)=
   \begin{cases}
   p^{m},&\qquad\text{if}\ a^{(q-1)/(p^{d}+1)}\ne (-1)^{m/d}\ \text{and}\ m/d\ \text{even},\\
   -p^{m},&\qquad\text{if}\ a^{(q-1)/(p^{d}+1)}\ne (-1)^{m/d}\ \text{and}\ m/d\ \text{odd},\\
   p^{m+d},&\qquad\text{if}\ a^{(q-1)/(p^{d}+1)}= (-1)^{m/d}\ \text{and}\ m/d\ \text{odd},\\
   -p^{m+d},&\qquad\text{if}\ a^{(q-1)/(p^{d}+1)}= (-1)^{m/d}\ \text{and}\ m/d\ \text{even}.
   \end{cases}
  \end{equation*}
\end{sec2_lemma8}  

\section{Main results and their proofs}
This section is devoted to the presentation of  the main results of this paper, and their proofs by using the series of lemmas of the previous section. Lemma \ref{s3_lemma1_Aa} and \ref{s3_lemma2_bac} compute the  exponential sums (\ref{s1_Aa_definition}) and (\ref{s1_Bac_definition}), respectively; while Theorem \ref{s3_thm1} and \ref{s3_thm2_Nac}  explicitly give the set cardinalities of (\ref{s1_definingset}) and (\ref{s1_definition_Mac}).

\begin{sec3_lemma1}\label{s3_lemma1_Aa}
Let $ q=p^{e} $ where $ p $ is an odd prime and $ e $ a positive integer, $ a\in \mathbb{F}_{p}^{*} $, and $ d=\gcd(\alpha,e) $. Suppose $ \frac{e}{d} $ to be odd. Then, 
\begin{itemize}
\item if $ e $ is even with $ e=2m $, we have
\begin{equation*}
   A_{\alpha}(0)=
   \begin{cases}
   -(p-1)p^{m}&\quad\text{if}\ p\equiv 1\pmod 4,\\
   -(-1)^{m}(p-1)p^{m}&\quad\text{if}\ p\equiv 3\pmod 4,
    \end{cases}
\end{equation*}
and
\begin{equation*}
  \\ A_{\alpha}(a)=
   \begin{cases}
   p^{m}&\quad\text{if}\ p\equiv 1\pmod 4,\\
   (-1)^{m}p^{m}&\quad\text{if}\ p\equiv 3\pmod 4.
    \end{cases}
\end{equation*} 
\item If $ e $ is odd, then
\begin{equation*}
   A_{\alpha}(0)=
   \begin{cases}
   0&\quad\text{if}\ p\equiv 1\pmod 4,\\
   0&\quad\text{if}\ p\equiv 3\pmod 4,
    \end{cases}
\end{equation*}
and
\begin{equation*}
   A_{\alpha}(a)=
   \begin{cases}
  \widehat{\eta}(a)p^{\frac{e+1}{2}}&\quad\text{if}\ p\equiv 1\pmod 4,\\
   -\sqrt{-1}^{e+1}\widehat{\eta}(a)p^{\frac{e+1}{2}}&\quad\text{if}\ p\equiv 3\pmod 4.
    \end{cases}
\end{equation*} 
\end{itemize}
\end{sec3_lemma1}

\begin{IEEEproof}
From Lemma \ref{s2_lemma7_sa0_edodd} and (\ref{s1_Aa_definition}), it is easy to obtain that
\begin{equation}\label{s3_lemma_Aa_proof_eq1}
\begin{split}
A_{\alpha}(a)&=\sum_{y\in \mathbb{F}_{p}^{*}}\zeta_{p}^{-ay}\sum_{x\in \mathbb{F}_{q}}\zeta_{p}^{yTr(x^{p^{\alpha}+1})}
=\sum_{y\in \mathbb{F}_{p}^{*}}\zeta_{p}^{-ay}S_{\alpha}(y,0)\\
&=
\begin{cases}
\sum_{y\in \mathbb{F}_{p}^{*}}\zeta_{p}^{-ay}(-1)^{e-1}\sqrt{q}\eta(y)&\quad \text{if}\ p\equiv 1\pmod 4,\\
\sum_{y\in \mathbb{F}_{p}^{*}}\zeta_{p}^{-ay}(-1)^{e-1}\sqrt{-1}^{e}\sqrt{q}\eta(y)&\quad \text{if}\ p\equiv 3\pmod 4.
\end{cases}
\end{split}
\end{equation}
\begin{itemize}
\item Let $ e $ be even with $ e=2m $. By Lemma \ref{s2_lemma2_quaChar}, $ \eta(y) =1$ for all $ y\in \mathbb{F}_{p}^{*} $. From (\ref{s3_lemma_Aa_proof_eq1}), we have
\begin{equation*}
A_{\alpha}(a)=
\begin{cases}
-p^{m}\sum_{y\in \mathbb{F}_{p}^{*}}\zeta_{p}^{-ay}&\quad \text{if}\ p\equiv 1\pmod 4,\\
-(-1)^{m}p^{m}\sum_{y\in \mathbb{F}_{p}^{*}}\zeta_{p}^{-ay}&\quad \text{if}\ p\equiv 3\pmod 4.
\end{cases}
\end{equation*}
If $ a\ne 0 $, by the orthogonal property of the additive character of $ \mathbb{F}_{q} $, $ \sum_{y\in \mathbb{F}_{p}^{*}}\zeta_{p}^{-ay}=-1 $. Hence,
\begin{equation*}
A_{\alpha}(a)=
\begin{cases}
p^{m}&\quad \text{if}\ p\equiv 1\pmod 4,\\
(-1)^{m}p^{m}&\quad \text{if}\ p\equiv 3\pmod 4.
\end{cases}
\end{equation*}
If $ a= 0 $, $ \sum_{y\in \mathbb{F}_{p}^{*}}\zeta_{p}^{-ay}=p-1 $. 
 Hence,
\begin{equation*}
A_{\alpha}(0)=
\begin{cases}
-p^{m}(p-1)&\quad \text{if}\ p\equiv 1\pmod 4,\\
-(-1)^{m}p^{m}(p-1)&\quad \text{if}\ p\equiv 3\pmod 4.
\end{cases}
\end{equation*}
\item  Let $ e $ be odd. By Lemma \ref{s2_lemma2_quaChar}, $ \eta(y) =\widehat{\eta}(y)$ for all $ y\in \mathbb{F}_{p}^{*} $. From (\ref{s3_lemma_Aa_proof_eq1}) for $ a\ne 0 $, we have
\begin{equation}\label{s3_lemma_Aa_proof_eq2}
A_{\alpha}(a)=
\begin{cases}
\widehat{\eta}(a)p^{\frac{e}{2}}\sum_{y\in \mathbb{F}_{p}^{*}}\zeta_{p}^{-ay}\widehat{\eta}(-ay)&\quad \text{if}\ p\equiv 1\pmod 4,\\
-\sqrt{-1}^{e}\widehat{\eta}(a)p^{\frac{e}{2}}\sum_{y\in \mathbb{F}_{p}^{*}}\zeta_{p}^{-ay}\widehat{\eta}(-ay)&\quad \text{if}\ p\equiv 3\pmod 4.
\end{cases}
\end{equation}
In (\ref{s3_lemma_Aa_proof_eq2}) is used the fact that $ \widehat{\eta}(-1) =1 $ if $ p\equiv 1 \pmod 4 $, and $ \widehat{\eta}(-1) =-1 $ if $ p\equiv 3 \pmod 4 $ \cite{Bib17}. It is clear that $ \sum_{y\in \mathbb{F}_{p}^{*}}\zeta_{p}^{-ay}\widehat{\eta}(-ay)=G(\widehat{\eta},\widehat{\chi}) $. By Lemma \ref{s2_lemma1_gauss_sum}, $ G(\widehat{\eta},\widehat{\chi})=\sqrt{p} $ if $ p\equiv 1 \pmod 4 $, and $ G(\widehat{\eta},\widehat{\chi})=\sqrt{-p} $ if $ p\equiv 3 \pmod 4 $. Combining so far the mentioned fact and identities, and (\ref{s3_lemma_Aa_proof_eq2}), we can obtain
\begin{equation*}
A_{\alpha}(a)=
\begin{cases}
\widehat{\eta}(a)p^{\frac{e+1}{2}}&\quad \text{if}\ p\equiv 1\pmod 4,\\
-\sqrt{-1}^{e+1}\widehat{\eta}(a)p^{\frac{e+1}{2}}&\quad \text{if}\ p\equiv 3\pmod 4.
\end{cases}
\end{equation*}

In (\ref{s3_lemma_Aa_proof_eq1}), set $ a=0 $ and $ \eta(y) =\widehat{\eta}(y)$. It is easy to see that $ A_{\alpha}(0)=0 $ for both $ p\equiv 1 \pmod 4 $ and $ p\equiv 3 \pmod 4 $, since $  \sum_{y\in \mathbb{F}_{p}^{*}}\widehat{\eta}(y)= 0$. 
\end{itemize}
\end{IEEEproof}

\begin{sec3_lemma2}\label{s3_lemma2_bac}
Let $ q=p^{e} $ where $ p $ is an odd prime and $ e $ a positive integer, 
%$ (a,c)\in \mathbb{F}_{p}^{*}\times \mathbb{F}_{p}^{*} $, 
and $ d=\gcd(\alpha,e) $. In addition, suppose $ \frac{e}{d} $ to be odd. For the given $ b\in   \mathbb{F}_{q}^{*}$, let $ \gamma $ be the unique solution of $ f(x)=x^{p^{2\alpha}}+x=-b^{p^{\alpha}} $.
%{\renewcommand{\labelitemi}{$\triangleright$}
%{\renewcommand{\labelitemii}{$\triangleright$}
%{\renewcommand{\labelitemiii}{$\triangleright$}
\begin{enumerate}[\hspace{0.5cm}\bfseries {Case} 1]
\item $ a=c=0 $.
\begin{itemize}
\item Let $ e $ be even with $ e=2m $.
\begin{itemize}
\item If $ p\equiv 1 \pmod 4 $, then
       \begin{equation*}
            B_{\alpha}(0,0)=
            \begin{cases}
             -p^{m}(p-1)^{2},&\quad \text{if}\ Tr(\gamma^{p^{\alpha}+1})=0,\\
               p^{m}(p-1),&\quad \text{if}\ Tr(\gamma^{p^{\alpha}+1})\ne 0.
            \end{cases}
           \end{equation*}
\item If $ p\equiv 3 \pmod 4 $, then
        \begin{equation*}
                B_{\alpha}(0,0)=
                \begin{cases}
            -(-1)^{m}p^{m}(p-1)^{2},&\quad \text{if}\ Tr(\gamma^{p^{\alpha}+1})=0,\\
              (-1)^{m}p^{m}(p-1),&\quad \text{if}\ Tr(\gamma^{p^{\alpha}+1})\ne 0.
                \end{cases}
               \end{equation*}
\end{itemize}
\item Let $ e $ be odd.
\begin{itemize}
\item If $ p\equiv 1 \pmod 4 $, then
         \begin{equation*}
                B_{\alpha}(0,0)=
                \begin{cases}
            0,&\quad \text{if}\ Tr(\gamma^{p^{\alpha}+1})=0,\\
               \widehat{\eta}(Tr(\gamma^{p^{\alpha}+1}))p^{\frac{e+1}{2}}(p-1),&\quad \text{if}\ Tr(\gamma^{p^{\alpha}+1})\ne 0.
                \end{cases}
               \end{equation*} 
\item If $ p\equiv 3 \pmod 4 $, then
         \begin{equation*}
                    B_{\alpha}(0,0)=
                    \begin{cases}
                0,&\quad \text{if}\ Tr(\gamma^{p^{\alpha}+1})=0,\\
                 -(\sqrt{-1})^{e+1}\widehat{\eta}(Tr(\gamma^{p^{\alpha}+1}))p^{\frac{e+1}{2}}(p-1),&\quad \text{if}\ Tr(\gamma^{p^{\alpha}+1})\ne 0.
                    \end{cases}
                   \end{equation*} 
\end{itemize}
\end{itemize}
\item $ a=0,c\ne 0 $.
\begin{itemize}
\item Let $ e $ be even with $ e=2m $.
\begin{itemize}
\item If $ p\equiv 1 \pmod 4 $, then
          \begin{equation*}
                      B_{\alpha}(0,c)=
                      \begin{cases}
                       p^{m}(p-1),&\quad \text{if}\ Tr(\gamma^{p^{\alpha}+1})=0,\\
                         -p^{m},&\quad \text{if}\ Tr(\gamma^{p^{\alpha}+1})\ne 0.
                      \end{cases}
                     \end{equation*}
\item If $ p\equiv 3 \pmod 4 $, then
        \begin{equation*}
                             B_{\alpha}(0,c)=
                             \begin{cases}
                              (-1)^{m}p^{m}(p-1),&\quad \text{if}\ Tr(\gamma^{p^{\alpha}+1})=0,\\
                                -(-1)^{m}p^{m},&\quad \text{if}\ Tr(\gamma^{p^{\alpha}+1})\ne 0.
                             \end{cases}
                            \end{equation*}
\end{itemize}
\item Let $ e $ be odd.
\begin{itemize}
\item If $ p\equiv 1 \pmod 4 $, then
   \begin{equation*}
                   B_{\alpha}(0,c)=
                   \begin{cases}
               0,&\quad \text{if}\ Tr(\gamma^{p^{\alpha}+1})=0,\\
                  -\widehat{\eta}(Tr(\gamma^{p^{\alpha}+1}))p^{\frac{e+1}{2}},&\quad \text{if}\ Tr(\gamma^{p^{\alpha}+1})\ne 0.
                   \end{cases}
                  \end{equation*} 
\item If $ p\equiv 3 \pmod 4 $, then
             \begin{equation*}
                  B_{\alpha}(0,c)=
                   \begin{cases}
               0,&\quad \text{if}\ Tr(\gamma^{p^{\alpha}+1})=0,\\
                  \sqrt{-1}^{e+1}\widehat{\eta}(Tr(\gamma^{p^{\alpha}+1}))p^{\frac{e+1}{2}},&\quad \text{if}\ Tr(\gamma^{p^{\alpha}+1})\ne 0.
                   \end{cases}
                  \end{equation*} 
\end{itemize}
\end{itemize}
\item $ a\ne 0, c=0 $.
\begin{itemize}
\item Let $ e $ be even with $ e=2m $.
\begin{itemize}
\item If $ p\equiv 1 \pmod 4 $, then
         \begin{equation*}
                      B_{\alpha}(a,0)=
                      \begin{cases}
                       p^{m}(p-1),&\quad \text{if}\ Tr(\gamma^{p^{\alpha}+1})=0,\\
                         -(1+p\widehat{\eta}(aTr(\gamma^{p^{\alpha}+1})))p^{m},&\quad \text{if}\ Tr(\gamma^{p^{\alpha}+1})\ne 0.
                      \end{cases}
        \end{equation*}
\item If $ p\equiv 3 \pmod 4 $, then
       \begin{equation*}
                             B_{\alpha}(a,0)=
                             \begin{cases}
                              (-1)^{m}p^{m}(p-1),&\quad \text{if}\ Tr(\gamma^{p^{\alpha}+1})=0,\\
                                -(-1)^{m}(1-p\widehat{\eta}(aTr(\gamma^{p^{\alpha}+1})))p^{m},&\quad \text{if}\ Tr(\gamma^{p^{\alpha}+1})\ne 0.
                             \end{cases}
               \end{equation*}
\end{itemize}
\item Let $ e $ be odd.
\begin{itemize}
\item If $ p\equiv 1 \pmod 4 $, then
     \begin{equation*}
                           B_{\alpha}(a,0)=
                           \begin{cases}
                            \widehat{\eta}(a)(p-1)p^{\frac{e+1}{2}},&\quad \text{if}\ Tr(\gamma^{p^{\alpha}+1})=0,\\
                          -(\widehat{\eta}(a)+\widehat{\eta}(Tr(\gamma^{p^{\alpha}+1})))p^{\frac{e+1}{2}},&\quad \text{if}\ Tr(\gamma^{p^{\alpha}+1})\ne 0.
                           \end{cases}
             \end{equation*}
\item If $ p\equiv 3 \pmod 4 $, then
\begin{equation*}
                           B_{\alpha}(a,0)=
                           \begin{cases}
                            -\sqrt{-1}^{e+1}\widehat{\eta}(a)(p-1)p^{\frac{e+1}{2}},&\quad \text{if}\ Tr(\gamma^{p^{\alpha}+1})=0,\\
                          \sqrt{-1}^{e+1}(\widehat{\eta}(a)+\widehat{\eta}(Tr(\gamma^{p^{\alpha}+1})))p^{\frac{e+1}{2}},&\quad \text{if}\ Tr(\gamma^{p^{\alpha}+1})\ne 0.
                           \end{cases}
             \end{equation*}
\end{itemize}
\end{itemize}
\item $ ac\ne 0 $.
\begin{itemize}
\item Let $ e $ be even with $ e=2m $.
\begin{itemize}
\item If $ p\equiv 1 \pmod 4 $, then
    \begin{equation*}
                         B_{\alpha}(a,c)=
                         \begin{cases}
                          -p^{m},&\quad \text{if}\ Tr(\gamma^{p^{\alpha}+1})=0,\\
                          -p^{m},&\quad \text{if}\ Tr(\gamma^{p^{\alpha}+1})=c^{2}/(4a),\\
                         -(1+p\widehat{\eta}(c^{2}-4aTr(\gamma^{p^{\alpha}+1})))p^{m},&\quad \text{otherwise}.
                         \end{cases}
           \end{equation*}
\item If $ p\equiv 3 \pmod 4 $, then
    \begin{equation*}
                            B_{\alpha}(a,c)=
                            \begin{cases}
                             -(-1)^{m}p^{m},&\quad \text{if}\ Tr(\gamma^{p^{\alpha}+1})=0,\\
                             -(-1)^{m}p^{m},&\quad \text{if}\ Tr(\gamma^{p^{\alpha}+1})=c^{2}/(4a),\\
                            -(-1)^{m}(1+p\widehat{\eta}(c^{2}-4aTr(\gamma^{p^{\alpha}+1})))p^{m},&\quad \text{otherwise}.
                            \end{cases}
              \end{equation*}
\end{itemize}
\item Let $ e $ be odd.
\begin{itemize}
\item If $ p\equiv 1 \pmod 4 $, then
     \begin{equation*}
                         B_{\alpha}(a,c)=
                           \begin{cases}
                                -\widehat{\eta}(a)p^{\frac{e+1}{2}},&\quad \text{if}\ Tr(\gamma^{p^{\alpha}+1})=0,\\
                           (\widehat{\eta}(Tr(\gamma^{p^{\alpha}+1}))(p-1)-\widehat{\eta}(a))p^{\frac{e+1}{2}},&\quad \text{if}\ Tr(\gamma^{p^{\alpha}+1})= c^{2}/(4a),\\
                              -(\widehat{\eta}(Tr(\gamma^{p^{\alpha}+1}))+\widehat{\eta}(a))p^{\frac{e+1}{2}},&\quad \text{otherwise}.
                               \end{cases}
                 \end{equation*}
\item If $ p\equiv 3 \pmod 4 $, then
\begin{equation*}
                         B_{\alpha}(a,c)=
                           \begin{cases}
                                \sqrt{-1}^{e+1}\widehat{\eta}(a)p^{\frac{e+1}{2}},&\quad \text{if}\ Tr(\gamma^{p^{\alpha}+1})=0,\\
                            -\sqrt{-1}^{e+1}(\widehat{\eta}(Tr(\gamma^{p^{\alpha}+1}))(p-1)-\widehat{\eta}(a))p^{\frac{e+1}{2}},&\quad \text{if}\ Tr(\gamma^{p^{\alpha}+1})= c^{2}/(4a),\\
                               \sqrt{-1}^{e+1}(\widehat{\eta}(Tr(\gamma^{p^{\alpha}+1}))+\widehat{\eta}(a))p^{\frac{e+1}{2}},&\quad \text{otherwise}.
                               \end{cases}
                 \end{equation*}
\end{itemize}
\end{itemize}
\end{enumerate}
\end{sec3_lemma2}
\begin{IEEEproof}
We only prove \textbf{Case} $ ac\ne 0 $ since the proofs of other cases are similar and easier. Since $ \frac{e}{d} $ is odd, $ f(x)=x^{p^{2\alpha}}+x=0 $ has no solution in $ \mathbb{F}_{q}^{*} $, and is a permutation polynomial over $ \mathbb{F}_{q} $ \cite{Bib9,Bib10,Bib16}. Hence, for the given $ b\in \mathbb{F}_{q}^{*} $, the equation $ f(x)=-b^{p^{\alpha}} $ has an unique solution in  $ \mathbb{F}_{q}^{*} $. Denote the unique solution as $ \gamma $, and let $ y,z\in \mathbb{F}_{p}^{*} $. It is easy to see that $ y^{-1}\gamma z $ is the unique solution of the equation $ y^{p^{\alpha}}x^{p^{2\alpha}}+yx=-(bz)^{p^{\alpha}}$. From (\ref{formula_sab_edodd}) and (\ref{s2_coulter_constant}), we have
\begin{equation}\label{s3_lemma2_bac_proof_eq01}
\begin{split}
S_{\alpha}(y,bz)&=\kappa\eta(-y)\overline{\chi(y(y^{-1}\gamma z)^{p^{\alpha}+1})}\\ 
&=\kappa\eta(-y)\overline{\chi(y^{-1}\gamma^{p^{\alpha}+1}z^{2})}\\ 
&=\kappa\eta(-y)\zeta_{p}^{-y^{-1}z^{2}Tr(\gamma^{p^{\alpha}+1})}.
\end{split}
\end{equation}
From the definition of $ B_{\alpha}(a,c) $ in (\ref{s1_Bac_definition}), and (\ref{s3_lemma2_bac_proof_eq01}), we have
\begin{equation}\label{s3_lemma2_bac_proof_eq02}
\begin{split}
B_{\alpha}(a,c)&=\sum_{y\in \mathbb{F}_{p}^{*}}\sum_{z\in \mathbb{F}_{p}^{*}}\zeta_{p}^{-ay-cz}\sum_{x\in \mathbb{F}_{q}}\chi(yx^{p^{\alpha}+1}+bzx)\\
&=\sum_{y\in \mathbb{F}_{p}^{*}}\sum_{z\in \mathbb{F}_{p}^{*}}\zeta_{p}^{-ay-cz}S_{\alpha}(y,bz)\\
&=\sum_{y\in \mathbb{F}_{p}^{*}}\sum_{z\in \mathbb{F}_{p}^{*}}\zeta_{p}^{-ay-cz}\kappa\eta(-y)\zeta_{p}^{-y^{-1}z^{2}Tr(\gamma^{p^{\alpha}+1})}\\
&=\kappa\sum_{y\in \mathbb{F}_{p}^{*}}\eta(-y)\zeta_{p}^{-ay}\sum_{z\in \mathbb{F}_{p}^{*}}\zeta_{p}^{-y^{-1}Tr(\gamma^{p^{\alpha}+1})z^{2}-cz}.
%&=\kappa\sum_{y\in \mathbb{F}_{p}^{*}}\eta(-y)\zeta_{p}^{-ay}(\sum_{z\in \mathbb{F}_{p}}\zeta_{p}^{-y^{-1}Tr(\gamma^{p^{\alpha}+1})z^{2}-cz}-1)
\end{split}
\end{equation}

If $ Tr(\gamma^{p^{\alpha}+1})=0 $, then, from the last equality of (\ref{s3_lemma2_bac_proof_eq02}), we have
\begin{equation}\label{s3_lemma2_bac_proof_eq03}
\begin{split}
B_{\alpha}(a,c)&=\kappa\sum_{y\in \mathbb{F}_{p}^{*}}\eta(-y)\zeta_{p}^{-ay}\sum_{z\in \mathbb{F}_{p}^{*}}\zeta_{p}^{-cz}\\
&=-\kappa\sum_{y\in \mathbb{F}_{p}^{*}}\eta(-y)\zeta_{p}^{-ay}.
\end{split}
\end{equation}

Recall that the Gauss sum over $ \mathbb{F}_{p}^{*} $, $ G(\widehat{\eta},\widehat{\chi}) $, is equal to $ \sqrt{p} $ if $ p\equiv 1\pmod 4 $, and $ \sqrt{-p} $ if $ p\equiv 3\pmod 4 $ (see Lemma \ref{s2_lemma1_gauss_sum}). The values of the constant occurring in the Coulter's formulae (\ref{formula_sab_edodd}) are explicitly listed in (\ref{s2_coulter_cst_eeven}) and (\ref{s2_coulter_cst_eodd}) to ease computation. 

For $ Tr(\gamma^{p^{\alpha}+1})=0 $, two cases are distinguished:
\begin{itemize}
\item Let $ e $ be even with $ e=2m $. From Lemma \ref{s2_lemma2_quaChar}, $ \eta(-y)=1  $. Thus, (\ref{s3_lemma2_bac_proof_eq03}) becomes 
\begin{equation}\label{s3_lemma2_bac_proof_eq04}
\begin{split}
B_{\alpha}(a,c)&=-\kappa\sum_{y\in \mathbb{F}_{p}^{*}}\zeta_{p}^{-ay}\\
&=\kappa\\
&=
\begin{cases}
-p^{m}&\qquad\text{if}\ p\equiv 1\pmod 4,\\
-(-1)^{m}p^{m}&\qquad\text{if}\ p\equiv 3\pmod 4.
\end{cases}
\end{split}
\end{equation}
\item Let $ e $ be odd. From Lemma \ref{s2_lemma2_quaChar}, $ \eta(-y)=\widehat{\eta}(-y)  $. Thus, (\ref{s3_lemma2_bac_proof_eq03}) becomes 
\begin{equation}\label{s3_lemma2_bac_proof_eq05}
\begin{split}
B_{\alpha}(a,c)&=\kappa\sum_{y\in \mathbb{F}_{p}^{*}}\eta(-y)\zeta_{p}^{-ay}\sum_{z\in \mathbb{F}_{p}^{*}}\zeta_{p}^{-cz}\\
&=-\kappa\sum_{y\in \mathbb{F}_{p}^{*}}\widehat{\eta}(-y)\zeta_{p}^{-ay}\\
&=-\kappa\widehat{\eta}(a)\sum_{y\in \mathbb{F}_{p}^{*}}\widehat{\eta}(-ay)\zeta_{p}^{-ay}\\
&=-\kappa\widehat{\eta}(a)G(\widehat{\eta},\widehat{\chi})\\
&=
\begin{cases}
-\widehat{\eta}(a)p^{\frac{e+1}{2}}&\qquad\text{if}\ p\equiv 1\pmod 4,\\
\sqrt{-1}^{e+1}\widehat{\eta}(a)p^{\frac{e+1}{2}}&\qquad\text{if}\ p\equiv 3\pmod 4.
\end{cases}
\end{split}
\end{equation}
\end{itemize}

Now, suppose that $ Tr(\gamma^{p^{\alpha}+1})\ne 0 $. Following the last equality of (\ref{s3_lemma2_bac_proof_eq02}), we have 

\begin{equation}\label{s3_lemma2_bac_proof_eq06}
\begin{split}
B_{\alpha}(a,c)&=\kappa\sum_{y\in \mathbb{F}_{p}^{*}}\eta(-y)\zeta_{p}^{-ay}\sum_{z\in \mathbb{F}_{p}^{*}}\zeta_{p}^{-y^{-1}Tr(\gamma^{p^{\alpha}+1})z^{2}-cz}\\
&=\kappa\sum_{y\in \mathbb{F}_{p}^{*}}\eta(-y)\zeta_{p}^{-ay}(\sum_{z\in \mathbb{F}_{p}}\zeta_{p}^{-y^{-1}Tr(\gamma^{p^{\alpha}+1})z^{2}-cz}-1).
\end{split}
\end{equation}

By applying Lemma \ref{s2_lemma2_quafunc} to the sum $ \sum_{z\in \mathbb{F}_{p}}\zeta_{p}^{-y^{-1}Tr(\gamma^{p^{\alpha}+1})z^{2}-cz} $ in (\ref{s3_lemma2_bac_proof_eq06}) with $ a_{2}=-y^{-1}Tr(\gamma^{p^{\alpha}+1}), a_{1}=-c $, and $ a_{0}=0 $, we obtain
 
\begin{equation}\label{s3_lemma2_bac_proof_eq07}
\begin{split}
&\sum_{z\in \mathbb{F}_{p}}\zeta_{p}^{-y^{-1}Tr(\gamma^{p^{\alpha}+1})z^{2}-cz}\\
&=\sum_{z\in \mathbb{F}_{p}}\widehat{\chi}(-y^{-1}Tr(\gamma^{p^{\alpha}+1})z^{2}-cz)\\
&=\widehat{\chi}(\frac{c^{2}y}{4Tr(\gamma^{p^{\alpha}+1})})\widehat{\eta}(\frac{-Tr(\gamma^{p^{\alpha}+1})}{y})G(\widehat{\eta},\widehat{\chi})\\
&=\widehat{\eta}(Tr(\gamma^{p^{\alpha}+1}))G(\widehat{\eta},\widehat{\chi})\widehat{\eta}(-y)\zeta_{p}^{\frac{c^{2}y}{4Tr(\gamma^{p^{\alpha}+1})}}.
\end{split}
\end{equation}

Substitute the last equality of (\ref{s3_lemma2_bac_proof_eq07}) into (\ref{s3_lemma2_bac_proof_eq06}), we have

\begin{equation}\label{s3_lemma2_bac_proof_eq08}
\begin{split}
B_{\alpha}(a,c)&=\kappa\sum_{y\in \mathbb{F}_{p}^{*}}\eta(-y)\zeta_{p}^{-ay}(\widehat{\eta}(Tr(\gamma^{p^{\alpha}+1}))G(\widehat{\eta},\widehat{\chi})\widehat{\eta}(-y)\zeta_{p}^{\frac{c^{2}y}{4Tr(\gamma^{p^{\alpha}+1})}}-1)\\
&=\kappa\widehat{\eta}(Tr(\gamma^{p^{\alpha}+1}))G(\widehat{\eta},\widehat{\chi})\sum_{y\in \mathbb{F}_{p}^{*}}\eta(-y)\widehat{\eta}(-y)\zeta_{p}^{\frac{(c^{2}-4aTr(\gamma^{p^{\alpha}+1}))y}{4Tr(\gamma^{p^{\alpha}+1})}}-\kappa\sum_{y\in \mathbb{F}_{p}^{*}}\eta(-y)\zeta_{p}^{-ay}.\\
\end{split}
\end{equation}

Consider case $ Tr(\gamma^{p^{\alpha}+1})=c^{2}/(4a) $. Then, from (\ref{s3_lemma2_bac_proof_eq08}), we get 

\begin{equation}\label{s3_lemma2_bac_proof_eq09}
\begin{split}
B_{\alpha}(a,c)&=\kappa\widehat{\eta}(Tr(\gamma^{p^{\alpha}+1}))G(\widehat{\eta},\widehat{\chi})\sum_{y\in \mathbb{F}_{p}^{*}}\eta(-y)\widehat{\eta}(-y)-\kappa\sum_{y\in \mathbb{F}_{p}^{*}}\eta(-y)\zeta_{p}^{-ay}\\
\end{split}
\end{equation}

For $ Tr(\gamma^{p^{\alpha}+1})=c^{2}/(4a) $, two cases are distinguished:
\begin{itemize}
\item Let $ e $ be even with $ e=2m $. From Lemma \ref{s2_lemma2_quaChar}, $ \eta(-y)=1  $. Thus, (\ref{s3_lemma2_bac_proof_eq09}) becomes 
\begin{equation}\label{s3_lemma2_bac_proof_eq10}
\begin{split}
B_{\alpha}(a,c)&=\kappa\widehat{\eta}(Tr(\gamma^{p^{\alpha}+1}))G(\widehat{\eta},\widehat{\chi})\sum_{y\in \mathbb{F}_{p}^{*}}\widehat{\eta}(-y)-\kappa\sum_{y\in \mathbb{F}_{p}^{*}}\zeta_{p}^{-ay}\\
&=\kappa\widehat{\eta}(Tr(\gamma^{p^{\alpha}+1}))G(\widehat{\eta},\widehat{\chi})\cdot 0-\kappa\cdot (-1)\\
&=\kappa\\
&=
\begin{cases}
-p^{m}&\qquad\text{if}\ p\equiv 1\pmod 4,\\
-(-1)^{m}p^{m}&\qquad\text{if}\ p\equiv 3\pmod 4.
\end{cases}
\end{split}
\end{equation}
\item Let $ e $ be odd. From Lemma \ref{s2_lemma2_quaChar}, $ \eta(-y)=\widehat{\eta}(-y)  $. Thus, (\ref{s3_lemma2_bac_proof_eq09}) becomes 
\begin{equation}\label{s3_lemma2_bac_proof_eq11}
\begin{split}
B_{\alpha}(a,c)&=\kappa\widehat{\eta}(Tr(\gamma^{p^{\alpha}+1}))G(\widehat{\eta},\widehat{\chi})\sum_{y\in \mathbb{F}_{p}^{*}}\widehat{\eta}(-y)\widehat{\eta}(-y)-\kappa\sum_{y\in \mathbb{F}_{p}^{*}}\widehat{\eta}(-y)\zeta_{p}^{-ay}\\
&=\kappa\widehat{\eta}(Tr(\gamma^{p^{\alpha}+1}))G(\widehat{\eta},\widehat{\chi})\sum_{y\in \mathbb{F}_{p}^{*}}1-\kappa\widehat{\eta}(a)\sum_{y\in \mathbb{F}_{p}^{*}}\widehat{\eta}(-ay)\zeta_{p}^{-ay}\\
&=\kappa\widehat{\eta}(Tr(\gamma^{p^{\alpha}+1}))G(\widehat{\eta},\widehat{\chi})(p-1)-\kappa\widehat{\eta}(a)G(\widehat{\eta},\widehat{\chi})\\
&=\kappa G(\widehat{\eta},\widehat{\chi})(\widehat{\eta}(Tr(\gamma^{p^{\alpha}+1}))(p-1)-\widehat{\eta}(a))\\
&=
\begin{cases}
(\widehat{\eta}(Tr(\gamma^{p^{\alpha}+1}))(p-1)-\widehat{\eta}(a))p^{\frac{e+1}{2}}&\qquad\text{if}\ p\equiv 1\pmod 4,\\
-\sqrt{-1}^{e+1}(\widehat{\eta}(Tr(\gamma^{p^{\alpha}+1}))(p-1)-\widehat{\eta}(a))p^{\frac{e+1}{2}}&\qquad\text{if}\ p\equiv 3\pmod 4.
\end{cases}
\end{split}
\end{equation}
\end{itemize}

Now consider the most general case in (\ref{s3_lemma2_bac_proof_eq08}): $ Tr(\gamma^{p^{\alpha}+1})\ne 0 $ and $ Tr(\gamma^{p^{\alpha}+1})\ne c^{2}/(4a) $. Two cases must be dealt with:
\begin{itemize}
\item Let $ e $ be even with $ e=2m $. From Lemma \ref{s2_lemma2_quaChar}, $ \eta(-y)=1  $. Thus, (\ref{s3_lemma2_bac_proof_eq08}) becomes 
\begin{equation}\label{s3_lemma2_bac_proof_eq12}
\begin{split}
B_{\alpha}(a,c)&=\kappa\widehat{\eta}(Tr(\gamma^{p^{\alpha}+1}))G(\widehat{\eta},\widehat{\chi})\sum_{y\in \mathbb{F}_{p}^{*}}\widehat{\eta}(-y)\zeta_{p}^{\frac{(c^{2}-4aTr(\gamma^{p^{\alpha}+1}))y}{4Tr(\gamma^{p^{\alpha}+1})}}-\kappa\sum_{y\in \mathbb{F}_{p}^{*}}\zeta_{p}^{-ay}\\
&=\kappa\widehat{\eta}(-(c^{2}-4aTr(\gamma^{p^{\alpha}+1})))G(\widehat{\eta},\widehat{\chi})\sum_{y\in \mathbb{F}_{p}^{*}}\widehat{\eta}(\frac{(c^{2}-4aTr(\gamma^{p^{\alpha}+1}))y}{4Tr(\gamma^{p^{\alpha}+1})})\zeta_{p}^{\frac{(c^{2}-4aTr(\gamma^{p^{\alpha}+1}))y}{4Tr(\gamma^{p^{\alpha}+1})}}-\kappa\cdot (-1)\\
&=\kappa\widehat{\eta}(-(c^{2}-4aTr(\gamma^{p^{\alpha}+1})))G(\widehat{\eta},\widehat{\chi})^{2}+\kappa\\
&=\kappa(1+\widehat{\eta}(-(c^{2}-4aTr(\gamma^{p^{\alpha}+1})))G(\widehat{\eta},\widehat{\chi})^{2})\\
&=
\begin{cases}
-p^{m}(1+\widehat{\eta}(c^{2}-4aTr(\gamma^{p^{\alpha}+1}))p)&\qquad\text{if}\ p\equiv 1\pmod 4,\\
-(-1)^{m}p^{m}(1+\widehat{\eta}(c^{2}-4aTr(\gamma^{p^{\alpha}+1}))p)&\qquad\text{if}\ p\equiv 3\pmod 4.
\end{cases}
\end{split}
\end{equation}
\item Let $ e $ be odd. From Lemma \ref{s2_lemma2_quaChar}, $ \eta(-y)=\widehat{\eta}(-y)  $. Thus, (\ref{s3_lemma2_bac_proof_eq08}) becomes 
\begin{equation}\label{s3_lemma2_bac_proof_eq13}
\begin{split}
B_{\alpha}(a,c)&=\kappa\widehat{\eta}(Tr(\gamma^{p^{\alpha}+1}))G(\widehat{\eta},\widehat{\chi})\sum_{y\in \mathbb{F}_{p}^{*}}\widehat{\eta}(-y)^{2}\zeta_{p}^{\frac{(c^{2}-4aTr(\gamma^{p^{\alpha}+1}))y}{4Tr(\gamma^{p^{\alpha}+1})}}-\kappa\sum_{y\in \mathbb{F}_{p}^{*}}\widehat{\eta}(-y)\zeta_{p}^{-ay}\\
&=\kappa\widehat{\eta}(Tr(\gamma^{p^{\alpha}+1}))G(\widehat{\eta},\widehat{\chi})\sum_{y\in \mathbb{F}_{p}^{*}}\zeta_{p}^{\frac{(c^{2}-4aTr(\gamma^{p^{\alpha}+1}))y}{4Tr(\gamma^{p^{\alpha}+1})}}-\kappa\widehat{\eta}(a)\sum_{y\in \mathbb{F}_{p}^{*}}\widehat{\eta}(-ay)\zeta_{p}^{-ay}\\
&=\kappa\widehat{\eta}(Tr(\gamma^{p^{\alpha}+1}))G(\widehat{\eta},\widehat{\chi})\cdot (-1)-\kappa\widehat{\eta}(a)G(\widehat{\eta},\widehat{\chi})\\
&=-\kappa G(\widehat{\eta},\widehat{\chi})(\widehat{\eta}(Tr(\gamma^{p^{\alpha}+1}))+\widehat{\eta}(a))\\
&=
\begin{cases}
-(\widehat{\eta}(Tr(\gamma^{p^{\alpha}+1}))+\widehat{\eta}(a))p^{\frac{e+1}{2}}&\qquad\text{if}\ p\equiv 1\pmod 4,\\
\sqrt{-1}^{e+1}(\widehat{\eta}(Tr(\gamma^{p^{\alpha}+1}))+\widehat{\eta}(a))p^{\frac{e+1}{2}}&\qquad\text{if}\ p\equiv 3\pmod 4.
\end{cases}
\end{split}
\end{equation}
\end{itemize}
The proof is completed by gathering the results of (\ref{s3_lemma2_bac_proof_eq04})-(\ref{s3_lemma2_bac_proof_eq05}) and (\ref{s3_lemma2_bac_proof_eq10})-(\ref{s3_lemma2_bac_proof_eq13}).
\end{IEEEproof}

% Define 
%      \begin{equation}\label{codewords_length}
%      n_{\alpha}(a)=
%      \begin{cases}
%      | D_{\alpha}(a) \cup \lbrace 0\rbrace | &\qquad\text{if}\ a=0,\\
%      | D_{\alpha}(a)|,&\qquad\text{if}\ a\ne 0.
%      \end{cases}
%      \end{equation}
      
%      It is clear that $ n_{\alpha}(a) $  gives the length of codewords of the linear codes $ \mathcal{C}_{D_{\alpha}}(a,c) $ of (\ref{wang_LD}),  whose values are explicitly computed in  the following theorem:

The following theorem determines the cardinality of the subset (\ref{s1_definingset}), i.e., $ n_{\alpha}(a) $ of (\ref{codewords_length}):
      
\begin{sec3_thm1}\label{s3_thm1}
Let $ q=p^{e} $ where $ p $ is an odd prime and $ e $ a positive integer, 
%$ (a,c)\in \mathbb{F}_{p}^{*}\times \mathbb{F}_{p}^{*} $, 
and $ d=\gcd(\alpha,e) $. In addition, suppose $ \frac{e}{d} $ to be odd. Then, 
%the length of codewords, 
$ n_{\alpha}(a) $ with $ a\in \mathbb{F}_{p} $ of (\ref{codewords_length}), can be determined as follows:
\begin{itemize}
\item Let $ e $ be even with $ e=2m $.
\begin{itemize}
\item If $ p\equiv 1\pmod 4 $, then
   \begin{equation*}
   \begin{cases}
   n_{\alpha}(0)=p^{e-1}-(p-1)p^{m-1},\\
   n_{\alpha}(a)= p^{e-1}+p^{m-1}.
   \end{cases}
   \end{equation*}
   \item If $ p\equiv 3\pmod 4 $, then
   \begin{equation*}
      \begin{cases}
      n_{\alpha}(0)&=p^{e-1}-(-1)^{m}(p-1)p^{m-1},\\
      n_{\alpha}(a)&= p^{e-1}+(-1)^{m}p^{m-1}.
      \end{cases}
   \end{equation*}
\end{itemize}
 
\item Let $ e $ be odd.
\begin{itemize}
 \item If $ p\equiv 1\pmod 4 $, then
     \begin{equation*}
     \begin{cases}
     n_{\alpha}(0)=p^{e-1},\\
     n_{\alpha}(a)= p^{e-1}+\widehat{\eta}(a)p^{\frac{e-1}{2}}.
     \end{cases}
     \end{equation*}
     \item If $ p\equiv 3\pmod 4 $, then
     \begin{equation*}
        \begin{cases}
        n_{\alpha}(0)&=p^{e-1},\\
       n_{\alpha}(a)&=p^{e-1}-(\sqrt{-1})^{e+1}\widehat{\eta}(a)p^{\frac{e-1}{2}}.
        \end{cases}
     \end{equation*}
\end{itemize}
\end{itemize}
Where $ a\in \mathbb{F}_{p}^{*} $.
\end{sec3_thm1} 
\begin{IEEEproof}
From (\ref{codewords_length}), we have
\begin{equation}\label{s3_proof_na}
\begin{split}
 n_{\alpha}(a)&=\frac{1}{p}\sum_{x\in \mathbb{F}_{q} }\biggl(\sum_{y\in\mathbb{F}_{p}}\zeta_{p}^{y(Tr(x^{p^{\alpha}+1})-a)}\biggr)\\
&=p^{e-1}+\frac{1}{p}\sum_{y\in\mathbb{F}_{p}^{*}}\zeta_{p}^{-ay}\sum_{x\in \mathbb{F}_{q}}\zeta_{p}^{yTr(x^{p^{\alpha}+1})}\\
&=p^{e-1}+p^{-1}A_{\alpha}(a).
\end{split}
\end{equation}
The actual theorem can be easily proven by combining (\ref{s3_proof_na}) and the formulae of $ A_{\alpha}(a) $ in Lemma \ref{s3_lemma1_Aa}.
\end{IEEEproof}

% The next exponential sum over $ \mathbb{F}_{q} $ was defined in \cite{Bib7} in order to compute the Hamming weights of the code $ \mathcal{C}_{D_{\alpha}}(a,c) $ of (\ref{wang_LD}): 
%%    with the \textit{defining set} $ D_{a} $ specified in (\ref{defining_set_a}).
%     \begin{equation}\label{Nac_set}
%     N_{b}(a,c)=\lbrace x\in \mathbb{F}_{q}: Tr(x^{p^{\alpha}+1})=a\ \text{and}\ Tr(bx)+c=0\rbrace.
%     \end{equation}
%     Let $ wt(\mathbf{c}_{b}) $ denote the Hamming weight of the codeword 
%     \begin{equation}\label{cb_def}
%    \mathbf{c}_{b}=(Tr(bd_{1})+c,\cdots,Tr(bd_{n_{\alpha}(a)})+c), b\in \mathbb{F}_{q}^{*},
%     \end{equation}
%    where $ d_{i}\in D_{\alpha}(a), 1\leq i\leq n_{\alpha}(a) $.
%%      of the code $ \mathcal{C}_{D_{a}} $. 
%      It is easy to see that
%     \begin{equation}\label{wt_formula}
%     wt(\mathbf{c}_{b})=n_{\alpha}(a)-|N_{b}(a,c)|.
%     \end{equation}

Next theorem lists the  formulae that compute the cardinality of the set (\ref{s1_definition_Mac}), i.e., $ N_{\alpha}(a,c) $ of (\ref{Mac_length}):

 \begin{sec3_thm2}\label{s3_thm2_Nac}
 Let $ q=p^{e} $ where $ p $ is an odd prime and $ e $ a positive integer, 
 %$ (a,c)\in \mathbb{F}_{p}^{*}\times \mathbb{F}_{p}^{*} $, 
 and $ d=\gcd(\alpha,e) $. In addition, suppose $ \frac{e}{d} $ to be odd. For the given $ b\in   \mathbb{F}_{q}^{*}$, let $ \gamma $ be the unique solution of $ f(x)=x^{p^{2\alpha}}+x=-b^{p^{\alpha}} $.
 \begin{enumerate}[\hspace{0.5cm}\bfseries {Case} 1]
 \item $ a=0,c=0 $.
 \begin{itemize}
 \item Let $ e $ be even with $ e=2m $.
   \begin{itemize}
   \item If $ p\equiv 1\pmod 4 $, then
   \begin{equation*}
   N_{\alpha}(0,0)=
   \begin{cases}
   p^{e-2}-(p-1)p^{m-1},&\qquad\text{if}\ Tr(\gamma^{p^{\alpha}+1})=0,\\
   p^{e-2},&\qquad\text{if}\ Tr(\gamma^{p^{\alpha}+1})\ne 0.
   \end{cases}
   \end{equation*}
   \item If $ p\equiv 3\pmod 4 $, then
   \begin{equation*}
   N_{\alpha}(0,0)=
   \begin{cases}
   p^{e-2}-(-1)^{m}(p-1)p^{m-1},&\qquad\text{if}\ Tr(\gamma^{p^{\alpha}+1})=0,\\
   p^{e-2},&\qquad\text{if}\ Tr(\gamma^{p^{\alpha}+1})\ne 0.
   \end{cases}
   \end{equation*}
   \end{itemize}
 \item Let $ e $ be odd.
  \begin{itemize}
  \item  If $ p\equiv 1\pmod 4 $, then
  \begin{equation*}
  N_{\alpha}(0,0)=
  \begin{cases}
  p^{e-2},&\qquad\text{if}\ Tr(\gamma^{p^{\alpha}+1})=0,\\
  p^{e-2}+\widehat{\eta}(Tr(\gamma^{p^{\alpha}+1}))(p-1)p^{\frac{e-3}{2}},&\qquad\text{if}\ Tr(\gamma^{p^{\alpha}+1})\ne 0.
  \end{cases}
  \end{equation*}
  \item If $ p\equiv 3\pmod 4 $, then
  \begin{equation*}
  N_{\alpha}(0,0)=
  \begin{cases}
  p^{e-2},&\qquad\text{if}\ Tr(\gamma^{p^{\alpha}+1})=0,\\
  p^{e-2}-(\sqrt{-1})^{e+1}\widehat{\eta}(Tr(\gamma^{p^{\alpha}+1}))(p-1)p^{\frac{e-3}{2}},&\qquad\text{if}\ Tr(\gamma^{p^{\alpha}+1})\ne 0.
  \end{cases}
  \end{equation*}
  \end{itemize}
 \end{itemize}
 \item $ a=0,c\ne 0 $.
 \begin{itemize}
  \item Let $ e $ be even with $ e=2m $.
     \begin{itemize}
        \item If $ p\equiv 1\pmod 4 $, then
        \begin{equation*}
        N_{\alpha}(0,c)=
        \begin{cases}
        p^{e-2},&\qquad\text{if}\ Tr(\gamma^{p^{\alpha}+1})=0,\\
        p^{e-2}-p^{m-1},&\qquad\text{if}\ Tr(\gamma^{p^{\alpha}+1})\ne 0.
        \end{cases}
        \end{equation*}
        \item If $ p\equiv 3\pmod 4 $, then
        \begin{equation*}
        N_{\alpha}(0,c)=
               \begin{cases}
               p^{e-2},&\qquad\text{if}\ Tr(\gamma^{p^{\alpha}+1})=0,\\
               p^{e-2}-(-1)^{m}p^{m-1},&\qquad\text{if}\ Tr(\gamma^{p^{\alpha}+1})\ne 0.
               \end{cases}
        \end{equation*}
        \end{itemize}
  \item Let $ e $ be odd.
   \begin{itemize}
    \item  If $ p\equiv 1\pmod 4 $, then
    \begin{equation*}
    N_{\alpha}(0,c)=
    \begin{cases}
    p^{e-2},&\qquad\text{if}\ Tr(\gamma^{p^{\alpha}+1})=0,\\
    p^{e-2}-\widehat{\eta}(Tr(\gamma^{p^{\alpha}+1}))p^{\frac{e-3}{2}},&\qquad\text{if}\ Tr(\gamma^{p^{\alpha}+1})\ne 0.
    \end{cases}
    \end{equation*}
    \item If $ p\equiv 3\pmod 4 $, then
    \begin{equation*}
    N_{\alpha}(0,c)=
    \begin{cases}
    p^{e-2},&\qquad\text{if}\ Tr(\gamma^{p^{\alpha}+1})=0,\\
    p^{e-2}+(\sqrt{-1})^{e+1}\widehat{\eta}(Tr(\gamma^{p^{\alpha}+1}))p^{\frac{e-3}{2}},&\qquad\text{if}\ Tr(\gamma^{p^{\alpha}+1})\ne 0.
    \end{cases}
    \end{equation*}
    \end{itemize}
  \end{itemize}
 \item $ a\ne 0,c=0 $.
 \begin{itemize}
  \item Let $ e $ be even with $ e=2m $.
           \begin{itemize}
                   \item If $ p\equiv 1\pmod 4 $, then
                   \begin{equation*}
                   N_{\alpha}(a,0)=
                   \begin{cases}
                   p^{e-2}+p^{m-1},&\qquad\text{if}\ Tr(\gamma^{p^{\alpha}+1})=0,\\
                   p^{e-2}-\widehat{\eta}(aTr(\gamma^{p^{\alpha}+1}))p^{m-1},&\qquad\text{if}\ Tr(\gamma^{p^{\alpha}+1})\ne 0.
                   \end{cases}
                   \end{equation*}
                   \item If $ p\equiv 3\pmod 4 $, then
                   \begin{equation*}
                   N_{\alpha}(a,0)=
                          \begin{cases}
                          p^{e-2}+(-1)^{m}p^{m-1},&\qquad\text{if}\ Tr(\gamma^{p^{\alpha}+1})=0,\\
                           p^{e-2}+(-1)^{m}\widehat{\eta}(aTr(\gamma^{p^{\alpha}+1}))p^{m-1},&\qquad\text{if}\ Tr(\gamma^{p^{\alpha}+1})\ne 0.
                          \end{cases}
                   \end{equation*}
                   \end{itemize}
  \item Let $ e $ be odd.
                    \begin{itemize}
                       \item  If $ p\equiv 1\pmod 4 $, then
                       \begin{equation*}
                       N_{\alpha}(a,0)=
                       \begin{cases}
                       p^{e-2}+\widehat{\eta}(a)p^{\frac{e-1}{2}},&\qquad\text{if}\ Tr(\gamma^{p^{\alpha}+1})=0,\\
                       p^{e-2}-\widehat{\eta}(Tr(\gamma^{p^{\alpha}+1}))p^{\frac{e-3}{2}},&\qquad\text{if}\ Tr(\gamma^{p^{\alpha}+1})\ne 0.
                       \end{cases}
                       \end{equation*}
                       \item If $ p\equiv 3\pmod 4 $, then
                       \begin{equation*}
                       N_{\alpha}(a,0)=
                       \begin{cases}
                       p^{e-2}-(\sqrt{-1})^{e+1}\widehat{\eta}(a)p^{\frac{e-1}{2}},&\qquad\text{if}\ Tr(\gamma^{p^{\alpha}+1})=0,\\
                       p^{e-2}+(\sqrt{-1})^{e+1}\widehat{\eta}(Tr(\gamma^{p^{\alpha}+1}))p^{\frac{e-3}{2}},&\qquad\text{if}\ Tr(\gamma^{p^{\alpha}+1})\ne 0.
                       \end{cases}
                       \end{equation*}
                       \end{itemize}
  \end{itemize}
 \item $ ac\ne 0 $.
 \begin{itemize}
  \item Let $ e $ be even with $ e=2m $.
              \begin{itemize}
                                \item If $ p\equiv 1\pmod 4 $, then
                                \begin{equation*}
                                N_{\alpha}(a,c)=
                                \begin{cases}
                                p^{e-2},&\qquad\text{if}\ Tr(\gamma^{p^{\alpha}+1})=0,\\
                                p^{e-2},&\qquad\text{if}\ Tr(\gamma^{p^{\alpha}+1})=c^{2}/(4a),\\
                                p^{e-2}-\widehat{\eta}(c^{2}-4aTr(\gamma^{p^{\alpha}+1}))p^{m-1},&\qquad\text{otherwise}.
                                \end{cases}
                                \end{equation*}
                                \item If $ p\equiv 3\pmod 4 $, then
                                \begin{equation*}
                                 N_{\alpha}(a,c)=
                                 \begin{cases}
                                   p^{e-2},&\qquad\text{if}\ Tr(\gamma^{p^{\alpha}+1})=0,\\
                                   p^{e-2},&\qquad\text{if}\ Tr(\gamma^{p^{\alpha}+1})=c^{2}/(4a),\\
                                   p^{e-2}-(-1)^{m}\widehat{\eta}(c^{2}-4aTr(\gamma^{p^{\alpha}+1}))p^{m-1},&\qquad\text{otherwise}.
                                 \end{cases}
                                \end{equation*}
                                \end{itemize}    
  \item Let $ e $ be odd.
                \begin{itemize}
                                                \item If $ p\equiv 1\pmod 4 $, then
                                                \begin{equation*}
                                                N_{\alpha}(a,c)=
                                                \begin{cases}
                                                p^{e-2},&\qquad\text{if}\ Tr(\gamma^{p^{\alpha}+1})=0,\\
          p^{e-2}+\widehat{\eta}(Tr(\gamma^{p^{\alpha}+1}))(p-1)p^{\frac{e-3}{2}},&\qquad\text{if}\ Tr(\gamma^{p^{\alpha}+1})=c^{2}/(4a),\\
          p^{e-2}-\widehat{\eta}(Tr(\gamma^{p^{\alpha}+1}))p^{\frac{e-3}{2}},&\qquad\text{otherwise}.
                                                \end{cases}
                                                \end{equation*}
                                                \item If $ p\equiv 3\pmod 4 $, then
                                                \begin{equation*}
                                                 N_{\alpha}(a,c)=
                                                   \begin{cases}
                                                 p^{e-2},&\qquad\text{if}\ Tr(\gamma^{p^{\alpha}+1})=0,\\
          p^{e-2}-(\sqrt{-1})^{e+1}\widehat{\eta}(Tr(\gamma^{p^{\alpha}+1}))(p-1)p^{\frac{e-3}{2}},&\qquad\text{if}\ Tr(\gamma^{p^{\alpha}+1})=c^{2}/(4a),\\
                    p^{e-2}+(\sqrt{-1})^{e+1}\widehat{\eta}(Tr(\gamma^{p^{\alpha}+1}))p^{\frac{e-3}{2}},&\qquad\text{otherwise}.
                                                   \end{cases}
                                                \end{equation*}
                                                \end{itemize}   
  \end{itemize}
 \end{enumerate}  
 \end{sec3_thm2}   
 
 \begin{IEEEproof} 
 From the definition of $ N_{\alpha}(a,c) $ in (\ref{Mac_length}), it is easy to see that
 \begin{equation}\label{s3_nac_proof}
 \begin{split}
 N_{\alpha}(a,c)&=p^{-2}\sum_{x\in \mathbb{F}_{q}}\biggl(\sum_{y\in \mathbb{F}_{p}}\zeta_{p}^{y(Tr(x^{p^{\alpha}+1})-a)}\biggr)\biggl(\sum_{z\in \mathbb{F}_{p}}\zeta_{p}^{z(Tr(bx)-c)}\biggr)\\
 &=p^{-2}\sum_{x\in \mathbb{F}_{q}}\biggl(1+\sum_{y\in \mathbb{F}_{p}^{*}}\zeta_{p}^{y(Tr(x^{p^{\alpha}+1})-a)}\biggr)\biggl(1+\sum_{z\in \mathbb{F}_{p}^{*}}\zeta_{p}^{z(Tr(bx)-c)}\biggr)\\
 &=p^{e-2}+p^{-2}\sum_{y\in \mathbb{F}_{p}^{*}}\sum_{x\in \mathbb{F}_{q}}\zeta_{p}^{y(Tr(x^{p^{\alpha}+1})-a)}+p^{-2}\sum_{z\in \mathbb{F}_{p}^{*}}\sum_{x\in \mathbb{F}_{q}}\zeta_{p}^{z(Tr(bx)-c)}\\
 &+p^{-2}\sum_{y\in \mathbb{F}_{p}^{*}}\sum_{z\in \mathbb{F}_{p}^{*}}\sum_{x\in \mathbb{F}_{q}}\zeta_{p}^{Tr(yx^{p^{\alpha+1}}+zbx)-ay-cz}\\
 &=p^{e-2}+p^{-2}A_{\alpha}(a)+p^{-2}B_{\alpha}(a,c).\\
 \end{split}
 \end{equation}
 The actual theorem can be proven by substituting the formulae of both $ A_{\alpha}(a) $ and $ B_{\alpha}(a,c)$ from Lemma \ref{s3_lemma1_Aa} and \ref{s3_lemma2_bac} into the last equality of (\ref{s3_nac_proof}).
 \end{IEEEproof}

\section{Concluding Remarks}
In this paper, two  exponential sums over finite field related to the Coulter's polynomial, and the cardinalities of two subsets of $ \mathbb{F}_{q} $, are settled for case $ \frac{e}{d} $ odd, that may have potential application in the construction of linear codes with few weights.

\vfill


\end{document}